\documentclass{aa}
\usepackage{graphicx}
\usepackage[varg]{txfonts}
\usepackage{color}

\begin{document} 

\title{Distinguishing freezing and thawing dark energy models through measurements of the fine-structure constant}

\author{J. M. A. Vilas Boas\inst{1}
\and
D. M. N. Magano\inst{2}
\and
C. J. A. P. Martins\inst{1,3}
\and
A. Barbecho\inst{4}
\and
C. Serrano\inst{5}}
\institute{Centro de Astrof\'{\i}sica da Universidade do Porto, Rua das Estrelas, 4150-762 Porto, Portugal\\
\email{up201403623@fc.up.pt,Carlos.Martins@astro.up.pt}
\and
Instituto de Telecomunica\c c\~oes, IST, Universidade de Lisboa, Av. Rovisco Pais 1, 1049-001 Lisboa, Portugal\\
\email{duartemagano@tecnico.ulisboa.pt}
\and 
Instituto de Astrof\'{\i}sica e Ci\^encias do Espa\c co, CAUP, Rua das Estrelas, 4150-762 Porto, Portugal
\and
ETSETB, Universitat Polit\`ecnica de Catalunya, C. Jordi Girona 1-3, 08034 Barcelona, Spain
\and
ESEIAAT, Universitat Polit\`ecnica de Catalunya, C. Miquel Vives 27-37, 08222 Terrassa, Spain
}
\date{Submitted \today}

\abstract
{Mapping the behaviour of dark energy is a pressing task for observational cosmology. Phenomenological classification divides dynamical dark energy models into freezing and thawing, depending on whether the dark energy equation of state is approaching or moving away from $w=p/\rho=-1$. Moreover, in realistic dynamical dark energy models the dynamical degree of freedom is expected to couple to the electromagnetic sector, leading to variations of the fine-structure constant $\alpha$. We discuss the feasibility of distinguishing between the freezing and thawing classes of models with current and forthcoming observational facilities and using a parametrisation of the dark energy equation of state, which can have either behaviour, introduced by Mukhanov  as fiducial paradigm. We illustrate how freezing and thawing models lead to different redshift dependencies of $\alpha$, and use a combination of current astrophysical observations and local experiments to constrain this class of models, improving the constraints on the key coupling parameter by more than a factor of two, despite considering a more extended parameter space than the one used in previous studies. We also briefly discuss the improvements expected from future facilities and comment on the practical limitations of this class of parametrisations. In particular, we show that sufficiently sensitive data can distinguish between freezing and thawing models, at least if one assumes that the relevant parameter space does not include phantom dark energy models.}

\keywords{Cosmology: theory -- Cosmology: observations -- Dark energy -- Methods: statistical -- Cosmological parameters}
\titlerunning{Distinguishing freezing and thawing dark energy models}
\authorrunning{Vilas Boas et al.}
\maketitle

\section{Introduction}

The discovery of cosmic acceleration, first inferred from measurements of the luminosity distance of type Ia supernovae in 1998 \citep{SN1,SN2}, opened up a new avenue for cosmological research. The most pressing task in this endeavour is to identify the source of this acceleration---the so-called Dark Energy---and understand its theoretical origin. A cosmological constant $\Lambda$ is the simplest explanation consistent with the currently available data, but the well-known fine-tuning problems associated with this solution imply that alternative scenarios should be sought and actively tested. A natural alternative explanation would involve scalar fields, an example of which is the Higgs field \citep{ATLAS,CMS}. Such cosmological scalar fields would lead to dynamical dark energy scenarios.

One of the simplest ways to classify dynamical dark energy models is to divide them into `freezing' and `thawing', depending on the behaviour of the dark energy equation of state $w(z)$ \citep{Caldwell}: qualitatively, in the former,  $w(z)$ is approaching a value of $-1$ today, while in the latter it is moving away from it. Recent work of \citet{Marsh} suggests that if one imposes physical priors, most realistic models are thawing. This should be contrasted to the analysis of \citet{Huterer}, who studied models selected on the basis of phase space priors and found the opposite result. We note that one difference between the two studies is that the latter started evolving the scalar fields at redshift $z=3$ instead of at early times. Meanwhile, recent observational constraints such as those from the Planck mission in combination with other probes \citep{Planck,Planck18} may be interpreted as slightly favouring thawing models, as pointed out in \citet{Linder}, though we note that the behaviour of dark energy is only well constrained at low redshifts. On the other hand, the recent work of \citet{Andersen} suggests that given current constraints, standard background probes such as type Ia supernovae (even at higher redshifts than currently available) are not ideal tools to discriminate between these models.

If dynamical scalar fields are indeed present, one naturally expects them to couple to the rest of the model, unless a yet-unknown symmetry is postulated to suppress these couplings \citep{Carroll}. In particular, a coupling of the field to the electromagnetic sector will lead to spacetime variations of the fine-structure constant $\alpha$; see \citet{UZAN} and \citet{ROPP} for recent reviews on this topic. Indeed there have been some recent indications of such a variation \citep{Webb}, at the relative level of variation of a few parts per million, though other measurements tightly constrain them \citep{LP1,LP3,Kotus}. Regardless of the outcome of these studies (i.e. whether they provide detections of variations or simply null results) the measurements have cosmological implications that go beyond the mere fundamental nature of the tests themselves. One of the goals of this work is to highlight some of these cosmological implications.

In what follows we make the `minimal' (most natural) assumption that the dynamical scalar field responsible for the dark energy also leads to a redshift dependence of $\alpha$. These models are `referred to as `Class I' in the terminology of \citet{ROPP}. Any varying $\alpha$ model not in Class I is denoted Class II, as in that case dark energy and $\alpha$ variations would be due to two different physical mechanisms. Our goal is to quantify the extent to which, in Class I models, astrophysical measurements of $\alpha $ --- which can already be made up to a redshift of $z=4$ and may in the future be done at significantly higher redshifts --- allow us to distinguish freezing and thawing models that would not be distinguishable using only cosmological observables. We note that there is at least one previous example of a class of dark energy models in which, given current constraints, the dark energy cannot be distinguished from a cosmological constant by any current or foreseeable background cosmology observables: that of rolling tachyon field models studied by \citet{Tachyons}.

Specifically we use a phenomenological model introduced by \citet{Mukhanov}  as fiducial parametrisation for dark energy; this model may display a freezing or thawing behaviour depending on the value of one of the model's free parameters. We first update the observational constraints on the model first obtained by \cite{AnaMarta2}, taking advantage of several recent, new local and astrophysical data. These include further measurements of $\alpha$ and direct tests of the weak equivalence principle. We then discuss how measurements of $\alpha$ expected from the next generation of high-resolution ultra-stable optical spectrographs such as ESPRESSO \citep{ESPRESSO,ESPRESSO2} and later ELT-HIRES \citep{EELT} will improve these constraints. In so doing we identify two main limitations of this parametrisation related to the fact that freezing models lead (other things being equal) to stronger $\alpha$ variations than thawing models and will therefore be more constrained by null results, and to the fact that a completely generic parameter space (including both canonical and phantom dark energy models) will lead to full degeneracies between model parameters. The first is an unavoidable consequence of the definition of freezing and thawing models, while the second can be removed by adopting a physical prior that excludes phantom models, and we show that, at least under this assumption, sufficiently sensitive data can distinguish between freezing and thawing models, as well as between Class II and Class I models. 

\section{\label{scalars}Dynamical dark energy models}

Dynamical dark energy models may be usefully classified on the basis of the dynamics of the underlying field in its potential \citep{Caldwell}. In  thawing models, the field was formerly frozen away from its minimum by Hubble damping (and therefore has an effective equation of state similar to that of a cosmological constant $w(z)\sim-1$) and recently started rolling down and accelerating towards this minimum. Conversely, in freezing models, the field was formerly rolling down a relatively steep potential towards its minimum but the onset of the accelerated expansion of the universe slows down the field dynamics and makes it decelerate. Some upper and lower bounds on the dynamics of the two classes of models in the field phase space were discussed in \citet{Caldwell}, although broader classes of models studied in \citet{Marsh} or \citet{Huterer} do not necessarily fulfil these bounds, especially in the thawing case. 

\begin{figure}
\centering
\includegraphics[width=9cm]{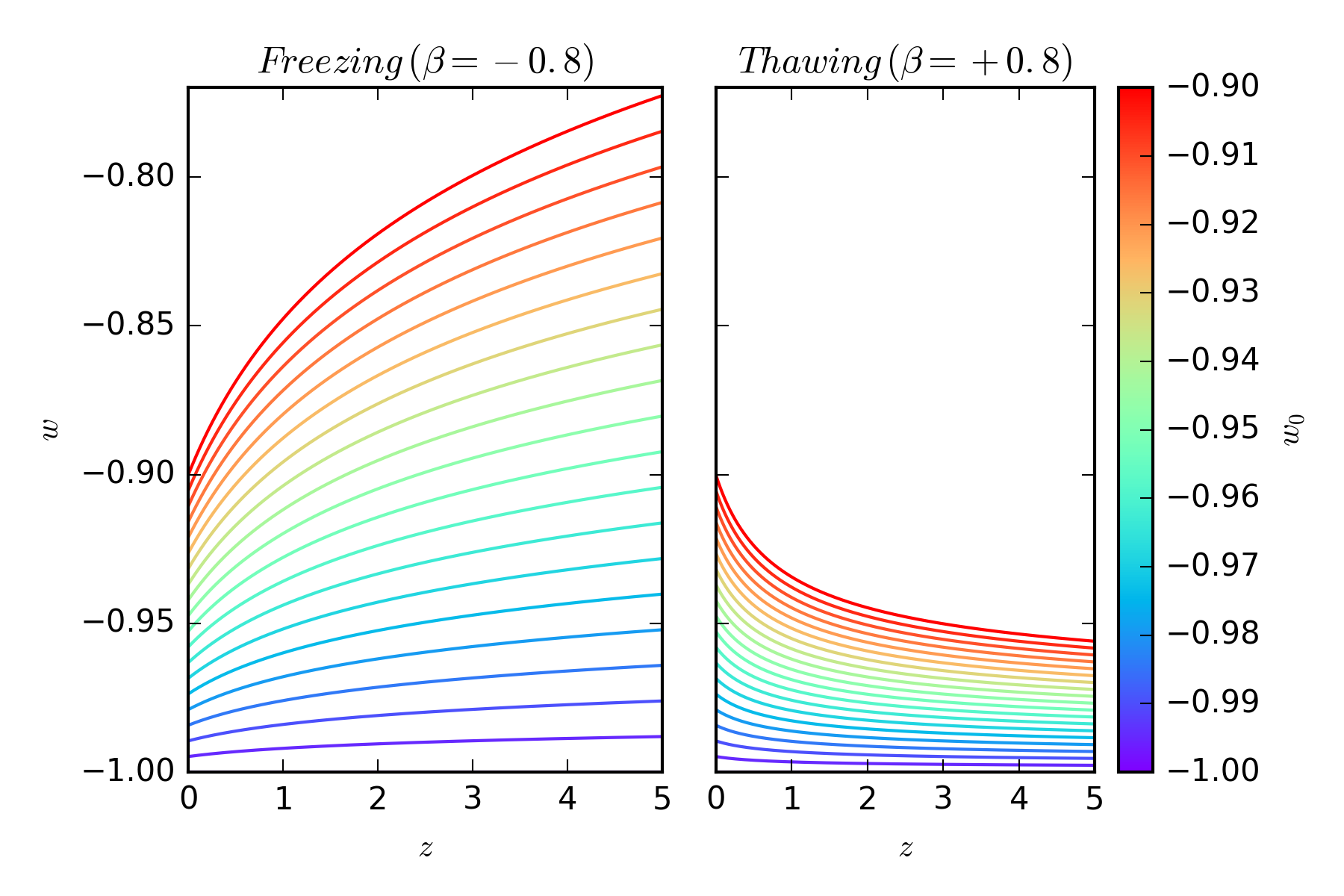}
\caption{Redshift dependence of the dark energy equation of state $w(z)$ in the parametrisation of \citet{Mukhanov}  for two specific cases of freezing ($\beta=-0.8$) and thawing ($\beta=0.8$) models, and for a range of values of $w_0$ identified by the colour bar.}
\label{fig1}%
\end{figure}

Dynamical scalar fields in an effective 4D field theory are naturally expected to couple to the rest of the theory, unless a (still unknown) symmetry is postulated to suppress this coupling \citep{Carroll,Dvali,Chiba}. In what follows we assume this to be the case for the dynamical degree of freedom responsible for the dark energy. Specifically we assume a coupling between the scalar field $\phi$ and the electromagnetic sector, which stems from a gauge kinetic function $B_F(\phi):$
\begin{equation}
{\cal L}_{\phi F} = - \frac{1}{4} B_F(\phi) F_{\mu\nu}F^{\mu\nu}\,.
\end{equation}
One can assume this function to be linear,
\begin{equation}
B_F(\phi) = 1- \zeta \kappa (\phi-\phi_0)\,,
\end{equation}
where $\kappa^2=8\pi G$, since, as pointed out in \citet{Dvali}, the absence of such a term would require the presence of a $\phi\to-\phi$ symmetry, but such a symmetry must be broken throughout most of the cosmological evolution. As is physically clear, the relevant parameter in the cosmological evolution is the field displacement relative to its present-day value, implying that we can set $\phi_0$  to zero without loss of generality. In these models the proton and neutron masses are also expected to vary due to the electromagnetic corrections of their masses; while we do not discuss this in detail in the present work --- detailed discussions can be found elsewhere \citep{UZAN,ROPP} --- one relevant consequence of this fact is that local tests of the equivalence principle also constrain these models.

A varying $\alpha$ violates the Einstein equivalence principle since it clearly violates local position invariance. It is also well known that varying fundamental couplings induce violations of the weak equivalence principle (or in other words, of the universality of free fall). The first detailed studies were done by \citet{Dicke}, and a more recent and thorough discussion can be found in \citet{Damour}. A light scalar field such as the one we are considering inevitably couples to nucleons due to the $\alpha$ dependence of their masses, and thus it mediates an isotope-dependent long-range force. This can be quantified through the dimensionless Eotvos parameter $\eta$. One can show that for the class of models we are considering the Eotvos parameter and the dimensionless coupling $\zeta$ are simply related by
\begin{equation}
\eta\sim10^{-3}\zeta^2\,;\label{eotvos}
\end{equation}
we use this relation to provide local constraints on our fiducial class of models.

\begin{figure}
\centering
\includegraphics[width=9cm]{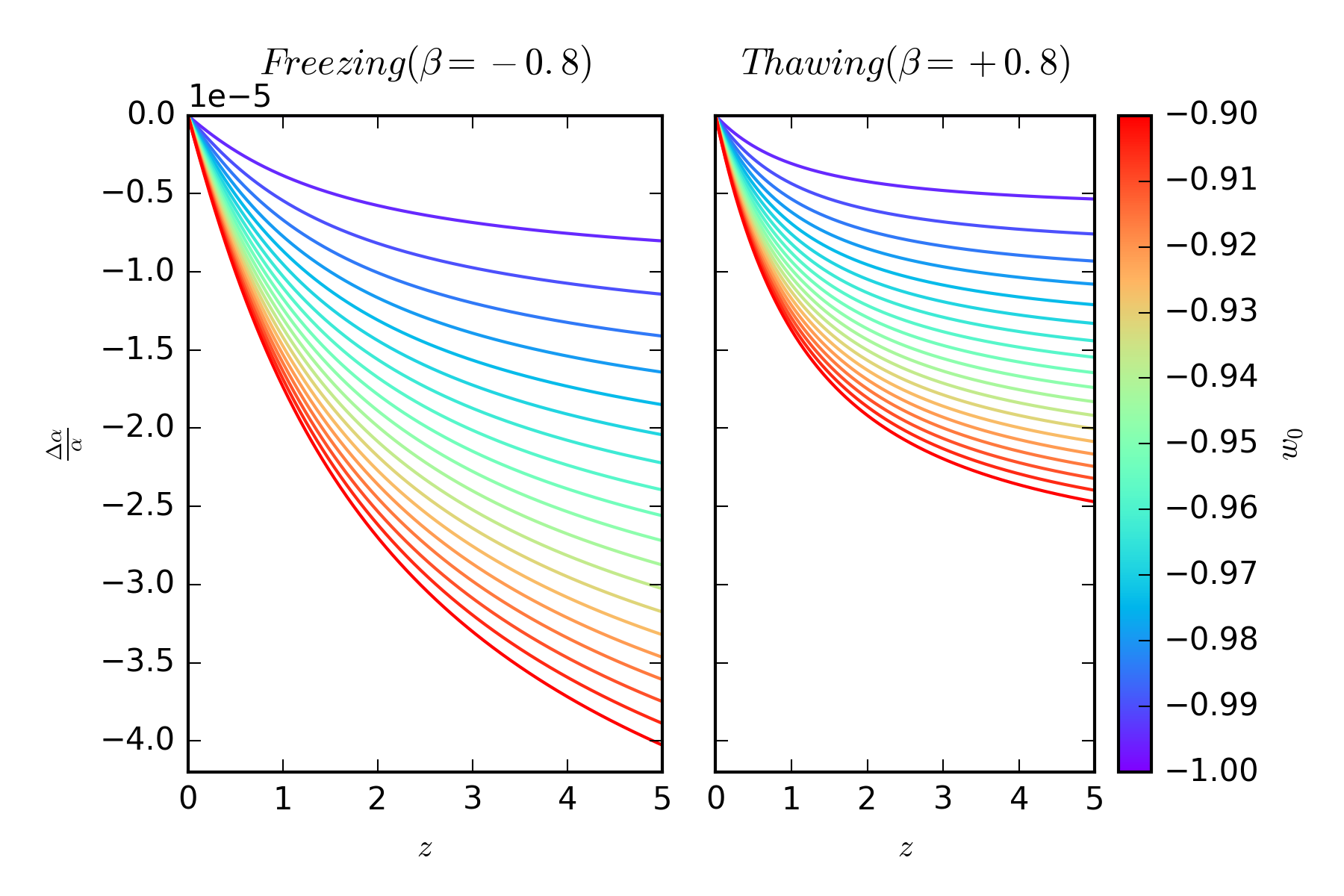}
\caption{Redshift dependence of the relative variation of $\alpha$ for different values of $w_0$, for freezing and thawing models. A value of the coupling $\zeta=-5\times10^{-6}$ has been assumed throughout.}
\label{fig2}%
\end{figure}
\begin{figure}
\centering
\includegraphics[width=9cm]{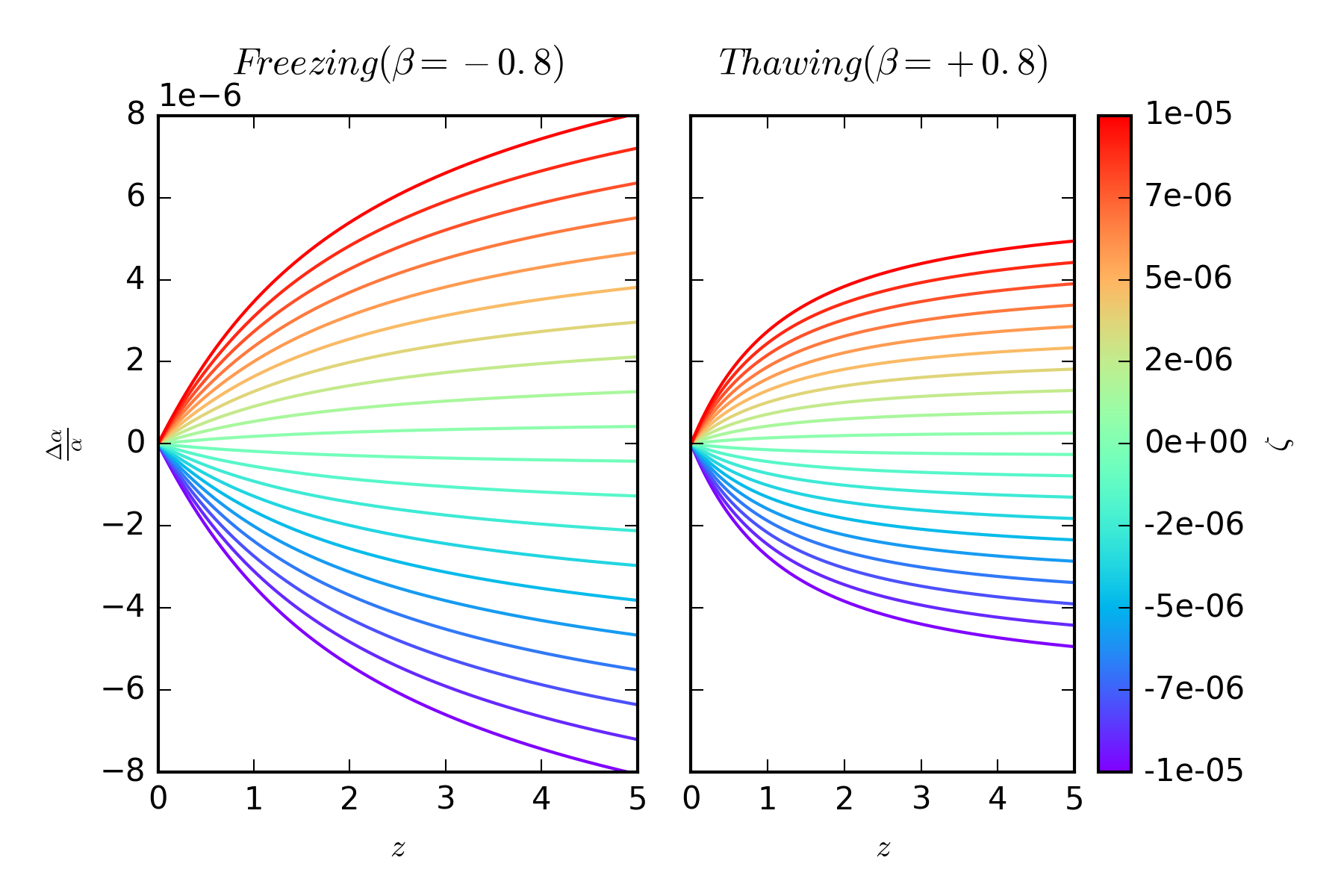}
\caption{Redshift dependence of the relative variation of $\alpha$ for different values of $\zeta$, for freezing and thawing models. A value of $w_0=-0.9$ has been assumed throughout.}
\label{fig3}%
\end{figure}

With these assumptions one can explicitly relate the evolution of $\alpha$ to that of dark energy, as in \citet{Erminia} and \citet{AnaMarta1} whose derivation we summarise here. The evolution of $\alpha$ can be written
\begin{equation}
\frac{\Delta \alpha}{\alpha} \equiv \frac{\alpha-\alpha_0}{\alpha_0} =B_F^{-1}(\phi)-1=
\zeta \kappa (\phi-\phi_0) \,,
\end{equation}
and defining the fraction of the dark energy density,
\begin{equation}
\Omega_\phi (z) \equiv \frac{\rho_\phi(z)}{\rho_{\rm tot}(z)} \simeq \frac{\rho_\phi(z)}{\rho_\phi(z)+\rho_m(z)} \,,
\end{equation}
where in the last step we neglect the contribution from radiation (since we are interested in low redshifts, $z<5$, where it is indeed negligible), the evolution of the putative scalar field can be expressed in terms of the dark energy properties $\Omega_\phi$ and $w$ as \citep{Nunes}
\begin{equation}
1+w_\phi = \frac{(\kappa\phi')^2}{3 \Omega_\phi} \,,
\end{equation}
with the prime denoting the derivative with respect to the logarithm of the scale factor. We finally obtain
\begin{equation} \label{eq:dalfa}
\frac{\Delta\alpha}{\alpha}(z) =\zeta \int_0^{z}\sqrt{3\Omega_\phi(z')\left(1+w_\phi(z')\right)}\frac{dz'}{1+z'}\,.
\end{equation}
We note that the above relation assumes a canonical scalar field. An analogous calculation for phantom fields \citep{Phantom} will lead to 
\begin{equation} \label{eq:dalfa2}
\frac{\Delta\alpha}{\alpha}(z) =-\zeta \int_0^{z}\sqrt{3\Omega_\phi(z')\left|1+w_\phi(z')\right|}\frac{dz'}{1+z'}\,;
\end{equation}
the change of sign stems from the fact that one expects phantom field to roll up the potential rather than down.

Since the redshift dependence of the dark energy equation of state $w(z)$ is qualitatively different in freezing and thawing models, the behaviour of $\alpha(z)$ will be correspondingly different in the two classes of models. This could in principle be observationally exploited: as we discuss below, mapping the behaviour of $\alpha(z)$ may be used to distinguish between the two classes of models, even allowing for the uncertainty in the coupling parameter $\zeta$ which provides an overall normalisation. Naturally, the detection of a redshift dependence of $\alpha$ immediately excludes $\Lambda$CDM, and thus distinguishes between a dynamical dark energy model and a cosmological constant.

\section{\label{sec:muk}Mukhanov's parametrisation}

Here we assume that the dark energy which is accelerating the universe can be phenomenologically described by a parametrisation proposed by \citet{Mukhanov}. This was introduced with the goal of describing the dynamics of inflationary models, but it is clear that it can also be applied to the case of the recent acceleration of the universe. In this scenario the dark energy equation of state has the following redshift dependence:
\begin{equation}
w(z)=-1+\frac{1+w_0}{\left[1+\ln{(1+z)}\right]^\beta}\,.
\end{equation}
Here, $w_0$ is the present-day value of the dark energy equation of state. The other free parameter is the slope $\beta$, which controls how fast the dark energy equation of state changes with redshift; $\beta=0$ corresponds to a constant equation of state while negative and positive values of $\beta$ correspond, respectively, to freezing and thawing models. This behaviour is illustrated in Figure \ref{fig1}.

In this case the dark energy density evolves as
\begin{equation}
\label{eq: sol_1}
\frac{\rho_{\phi}}{\rho_{\phi0}}=\exp{\left[3\frac{1+w_0}{1-\beta}\left([1+\ln(1+z)]^{1-\beta}-1\right)\right]}\,,
\end{equation}
provided one has $\beta\neq1$. For the particular case $\beta=1$ one instead has
\begin{equation}
\label{eq: sol_2}
\frac{\rho_{\phi}}{\rho_{\phi0}}=[1+\ln(1+z)]^{3(1+w_0)}\,.
\end{equation}
We further assume flat Friedmann-Lema\^{\i}tre-Robertson-Walker models, and neglect the contribution of the radiation density, since we are concerned with observational data at low redshifts. In this case the Friedmann equation is
\begin{equation}
\label{eq:Fried}
\frac{H^2(z)}{H^2_0}=\Omega_m(1+z)^3+\Omega_\phi\frac{\rho_\phi(z)}{\rho_{\phi0}}\,,
\end{equation}
where $\Omega_m+\Omega_\phi=1$.

Under the assumption that this phenomenological parametrisation describes the dynamics of some underlying scalar field, the redshift dependence of the fine-structure constant $\alpha$ is then given by Equation \ref{eq:dalfa}. Figures \ref{fig2} and \ref{fig3} illustrate how the behaviour of the relative variation of $\alpha$ depends on the Mukhanov model parameters and the coupling $\zeta$. These make it clear that, other parameters being equal, a freezing model will have larger variations of $\alpha$ than a thawing model. This raises the question as to how one may be able to observationally distinguish between them.

\begin{figure*}
\centering
\includegraphics[width=9cm]{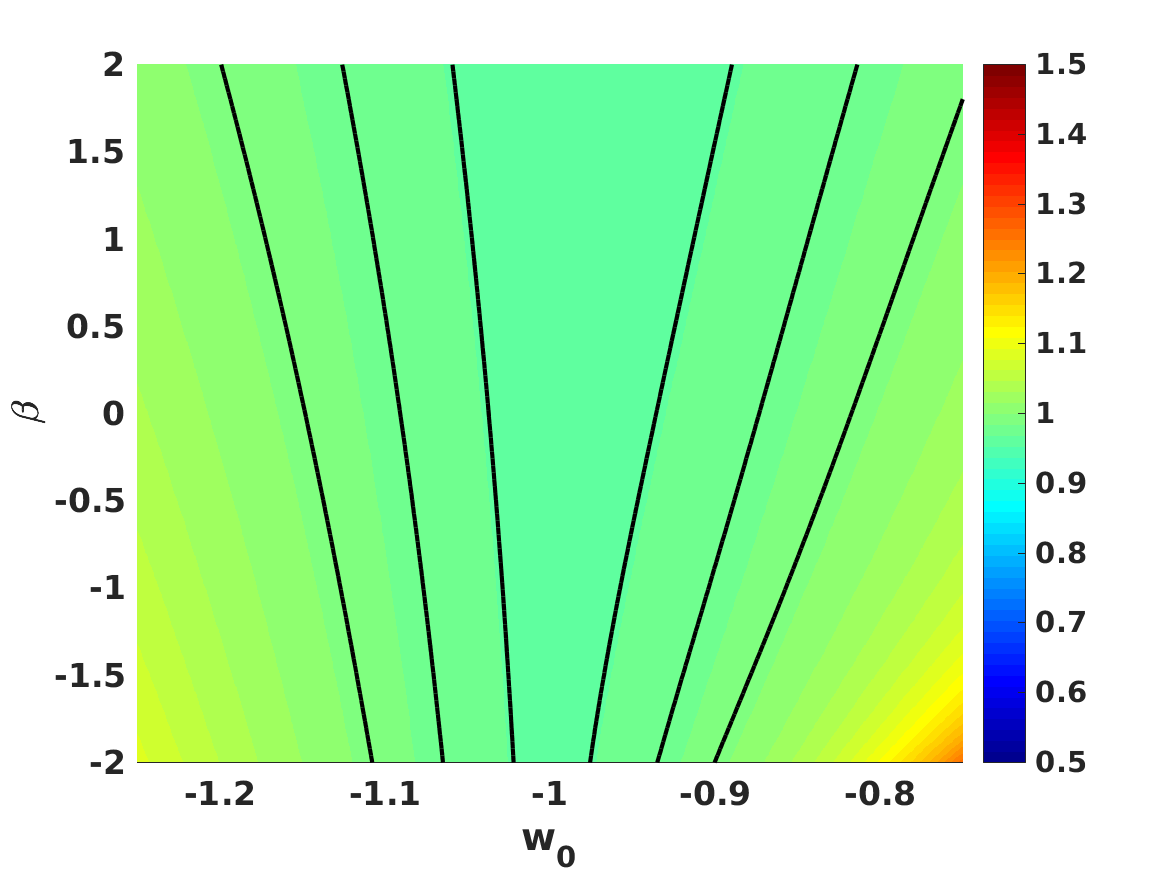}
\includegraphics[width=9cm]{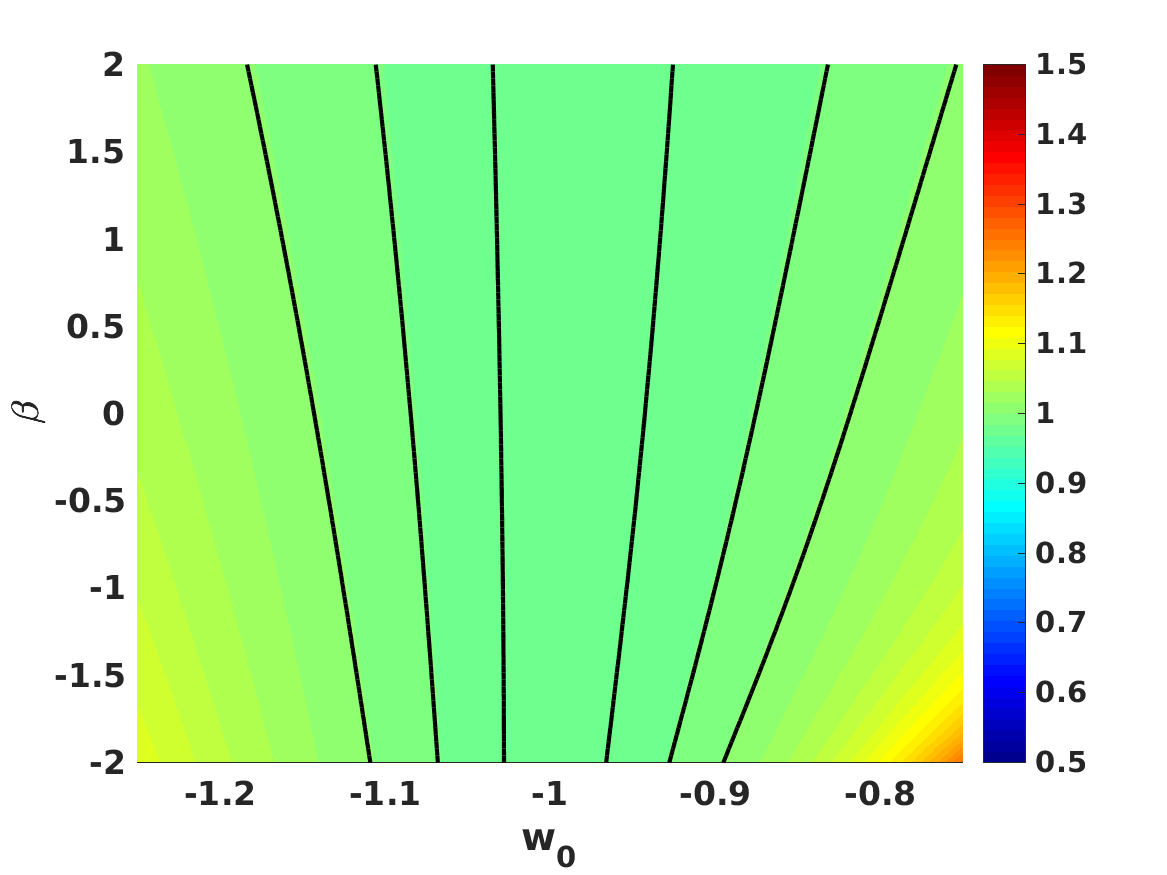}
\includegraphics[width=9cm]{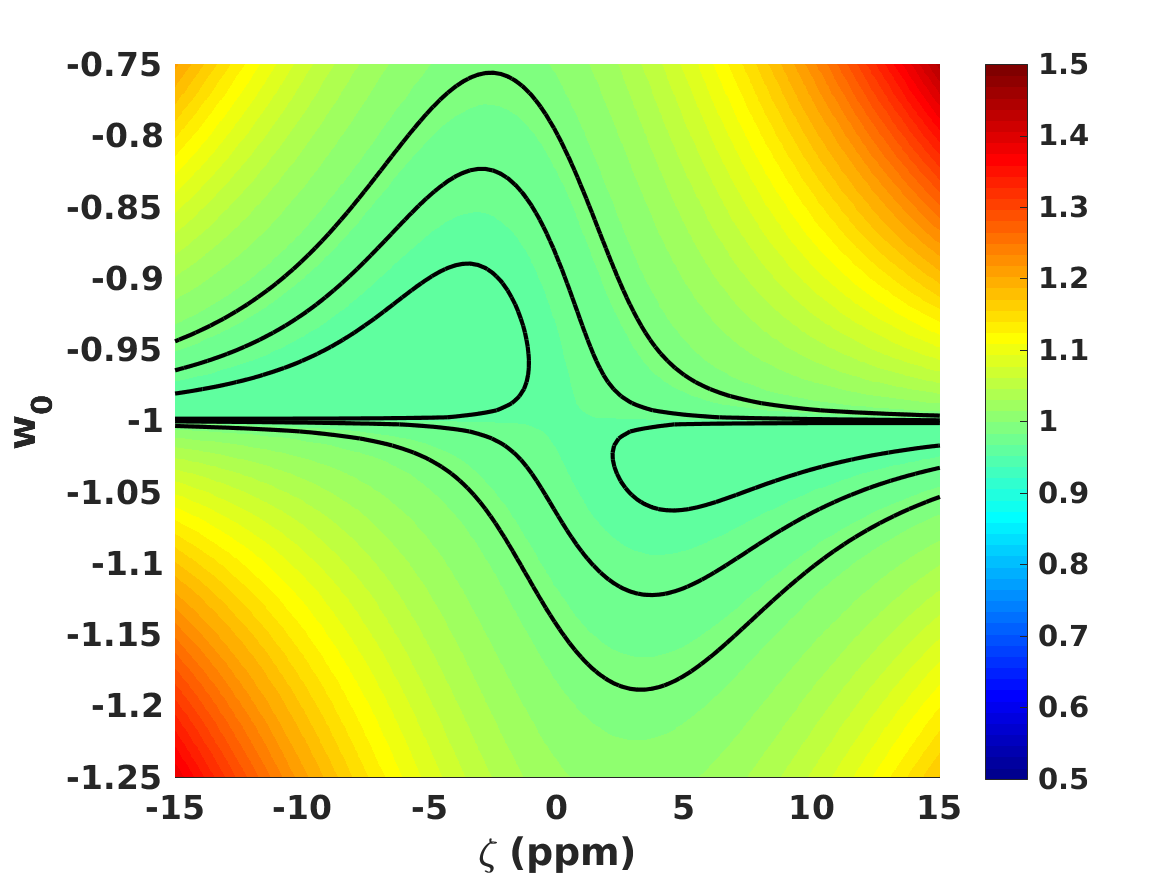}
\includegraphics[width=9cm]{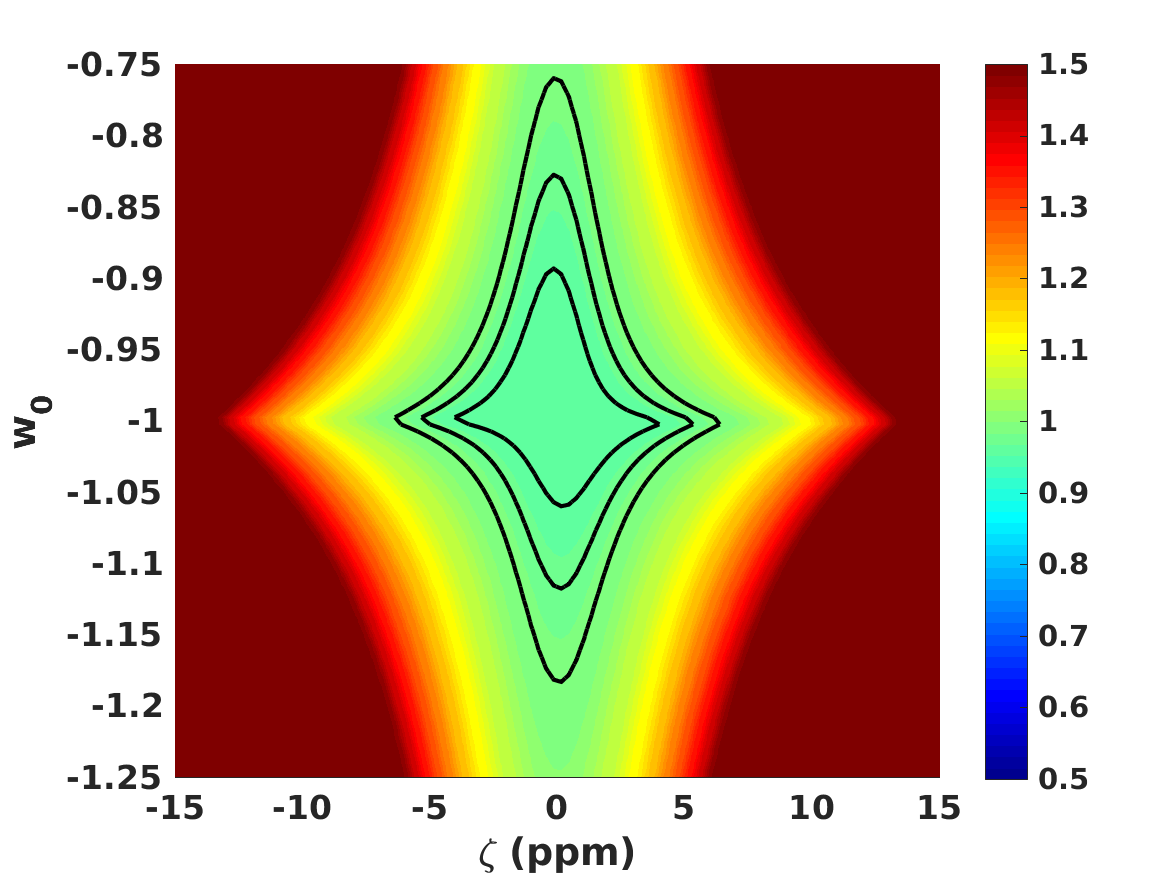}
\caption{Current constraints on dark energy models with the \citet{Mukhanov} dark energy equation of state, assuming a uniform prior on $w_0$. The left-hand side panels depict the constraints from the combination of astrophysical and cosmological constraints; in the right-hand side panels, local constraints from Oklo, atomic clocks, and MICROSCOPE have also been used. In all cases the black contours denote the one-, two-, and three-sigma confidence levels, and the colour map depicts the reduced chi-square of the fit for each set of model parameters (the dark red colour corresponds to a reduced chi-square of 1.5 or higher).}
\label{fig4}%
\end{figure*}
\begin{figure}[h!]
\centering
\includegraphics[width=9cm]{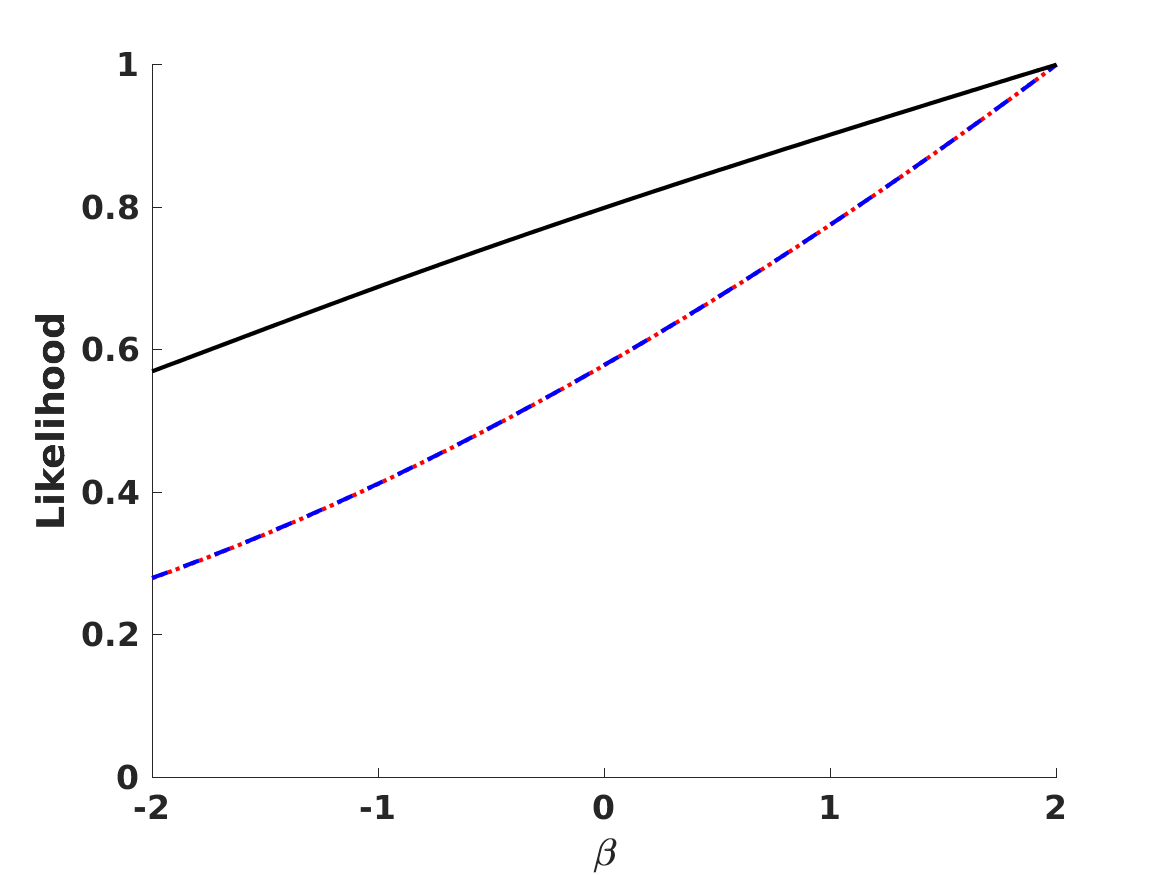}
\includegraphics[width=9cm]{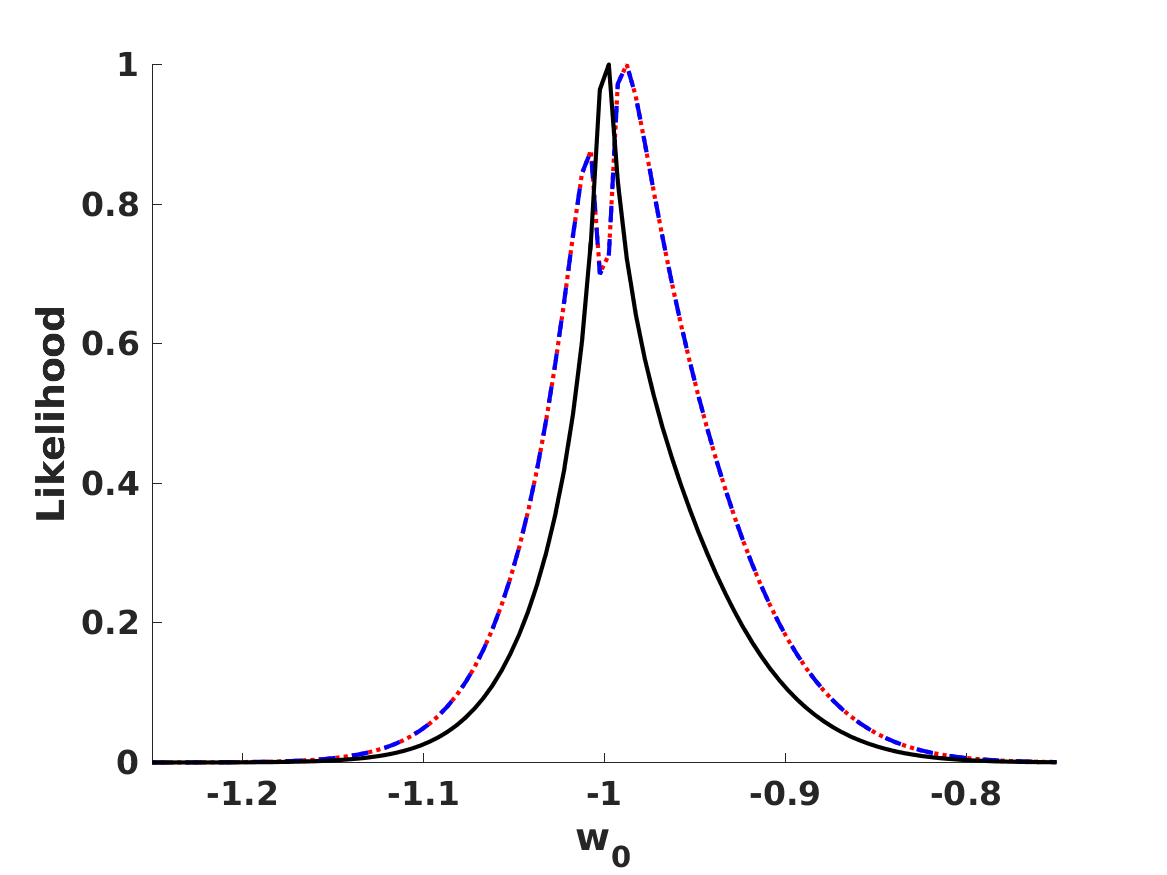}
\includegraphics[width=9cm]{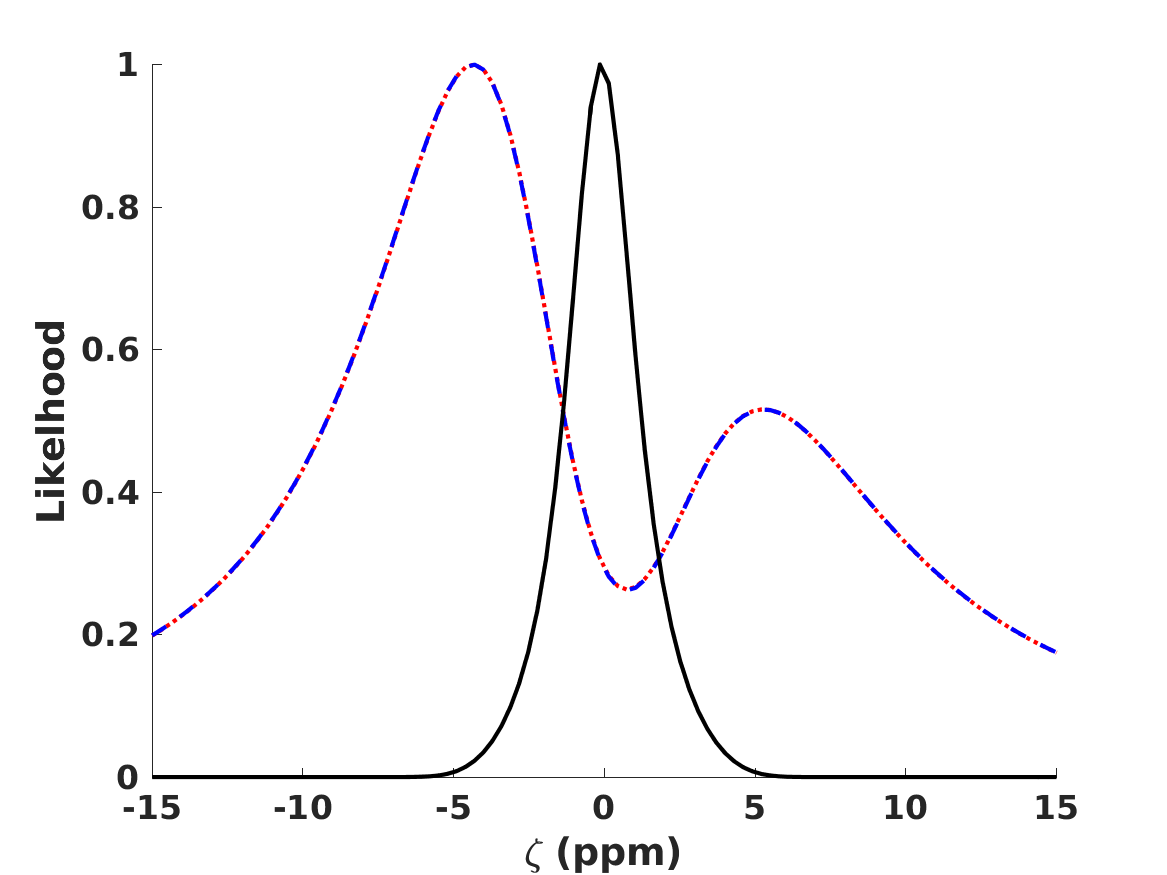}
\caption{Posterior likelihoods for the model parameters in the \citet{Mukhanov} dark energy equation of state, assuming a uniform prior on $w_0$ and marginalising the other parameters. The dot-dashed lines correspond to the constraints from the combination of astrophysical and cosmological constraints, while the solid lines correspond to constraints from our full data set (including also the local constraints from Oklo, atomic clocks and MICROSCOPE).}
\label{fig5}
\end{figure}

Finally, another important observable is the current drift rate of the value of $\alpha$, which from Equation \ref{eq:dalfa} can be found to be
\begin{equation}
\label{eq: drift}
\dfrac{1}{H_0} \dfrac{\dot{\alpha}}{\alpha} = -\zeta\sqrt{3 \Omega_{\phi 0} (1+w_0)}\,.
\end{equation}
Naturally this simply depends on the present value of the dark energy equation of state (and vanishes for $w_0=-1$), but it is independent of $\beta$. This provides a second way to constrain these models using local experiments.

\section{\label{current}Current observational constraints}

Below, we constrain this class of models using a combination of cosmological and astrophysical observations complemented by local tests. For background cosmology, we use the recent Pantheon catalogue of Type Ia supernovas of \citet{Riess}, including its covariance matrix, together with the compilation of 38 measurements of the Hubble parameter by \citet{Farooq}. The latter is important for extending the redshift lever arm and extending the overlap with the range of redshifts of $\alpha$ measurements. Specifically, this compilation includes Hubble parameter measurements up to redshift $z\sim2.36$, while the supernova data are all at effective redshifts $z<1.5$; for comparison, the $\alpha$ measurements in the following paragraph reach beyond $z=4$.

As for astrophysical (spectroscopic) tests of the stability of $\alpha$, we use both the \citet{Webb} data set (a large data set of 293 archival data measurements) and the smaller but more recent data set of 24 dedicated measurements \citep{ROPP,Cooksey}, which are expected to have a better control of possible systematic errors. The former data set spans a redshift range $0.22\le z \le 4.18$, while the latter spans a narrower range, $1.02\le z \le 2.13,$ but contains more stringent measurements that are compatible with the null result; overall these independent data sets complement each other and the constraining power of the two is comparable, as recently studied by \citet{Meritxell}.

Additionally we use three other local constraints. Firstly, the Oklo natural nuclear reactor \citep{Oklo} provides a geophysical constraint on the value of $\alpha$
\begin{equation} \label{okloalpha}
\frac{\Delta\alpha}{\alpha} =0.005\pm0.061\, ppm\,,
\end{equation}
at an effective redshift $z=0.14$, under the simplifying assumption that $\alpha$ is the only parameter that may have been different and all the remaining physics is unchanged. Secondly, the atomic clocks laboratory bound of \citet{Rosenband} constrains the current drift of $\alpha$ to be
\begin{equation} \label{rosen}
\left(\frac{1}{H}\frac{\dot\alpha}{\alpha}\right)_0=-0.22\pm0.32\, ppm\,;
\end{equation} 
we note that for convenience we use units of parts per million (ppm) throughout for values of $\alpha$ and the coupling $\zeta$. Thirdly, we use the recently improved constraint on the Eotvos parameter obtained by the MICROSCOPE satellite \citep{Touboul}:
\begin{equation} \label{micro}
\eta=(-0.1\pm1.3)\times10^{-14}\,.
\end{equation}
Compared to the previous study of this model in \citet{AnaMarta2}, there are therefore three main differences in these data sets:
\begin{itemize}
\item Here we use a more recent supernova data set; the previous work used the Union2.1 catalogue of \citet{Suzuki}
\item There are 10 additional measurements of the Hubble parameter and 13 additional measurements of $\alpha$; in both cases these are recent measurements which were not available at the time of the earlier analysis, highlighting the rapid progress in the field.
\item The Oklo and MICROSCOPE constraints were not previously used; the latter was also unavailable at the time of the early work.
\end{itemize}
As we see in what follows, the MICROSCOPE bound is the main contributor to the improvements in the current constraints on this class of models.

\begin{table}
\caption{Current two-sigma ($95.4\%$ c.l.) constraints for $w_0$ and $\zeta$, for the astrophysical and full data sets and for the uniform and logarithmic priors on $w_0$. For comparison, we also show the constraints previously obtained in \citet{AnaMarta2}. The reduced chi-square for the maximum likelihood parameters is also shown.}
\label{table1}
\centering
\begin{tabular}{| c | c | c || c |}
\hline
Parameter & Astro data & Full data & Previous \\
\hline
$\chi^2_\nu$  & 0.98 & 0.99 & 0.96 \\
$w_0$ & $-0.99^{+0.09}_{-0.08}$ & $-1.00^{+0.08}_{-0.06}$ & $-1.00^{+0.04}_{-0.03}$ \\
$\zeta$ & Unconstrained & $-0.1^{+2.7}_{-2.6}$ ppm & $0\pm6$ ppm \\
\hline
\end{tabular}
\end{table}

We carry out a standard statistical likelihood analysis, assuming uniform priors on the dark energy equation of state $w_0$, the slope $\beta,$ and the coupling $\zeta$. As previously mentioned we assume a flat universe, and further fix the present-day value of the matter density to $\Omega_m=0.3$, in agreement with both CMB and low-redshift data \citep{Planck18,Abbott:2017wau,Scolnic:2017caz,Jones:2017udy}. This choice is fully consistent with the supernova and Hubble parameter data we use, and does not significantly affect the parameters we are interested in constraining, especially the coupling $\zeta$; see \citet{Alves} for an analysis of this point in the case of different (but analogous) dark energy parametrisations. The Hubble constant is always analytically marginalised, following the prescription of \citet{Anagnostopoulos}.

Our constraints are summarised in Figures \ref{fig4} and \ref{fig5} and in Table \ref{table1}. The slope $\beta$ is not significantly constrained, though we also note an important asymmetry in the posterior likelihood for $\beta$: freezing models (having $\beta<0$) are more constrained than thawing models (with $\beta>0$). The reason for this was pointed out in \citet{AnaMarta2}: for a given value of the present-day dark energy equation of state (the other relevant dark energy parameter) a freezing model leads to larger variations of $\alpha$ at higher redshifts (assuming the same values of the coupling $\zeta$) and will therefore be more tightly constrained by astrophysical measurements. As the first panel in Figure \ref{fig5} shows, this asymmetry is reduced when the local data is added, since the atomic clocks and MICROSCOPE constraints (especially the latter) on the coupling are tight without being sensitive to $\beta$.

The bottom left panel of Figure \ref{fig4} and the bottom panel of Figure \ref{fig5} show that the combination of cosmological and astrophysical data has a one-sigma (thus not statistically significant) preference for a non-zero coupling $\zeta$. In fact, two distinct regions of parameter space are preferred: one has a negative coupling and a canonical dark energy equation of state, while the other has a positive coupling and a phantom dark energy equation of state. The reason for this degeneracy is that, as one can see from Eqs. \ref{eq:dalfa} and \ref{eq:dalfa2}, simultaneously changing the signs of $\zeta$ and of ($1+w_0$) leads to the same behaviour of $\alpha$. In any case, this preference disappears when the local constraints are added to the analysis.

In comparison to the previous analysis \citep{AnaMarta2}, the most significant improvement is that the constraint on the coupling $\zeta$ (which, in the absence of local data, is still unconstrained at two standard deviations) is improved by more than a factor of two. On the other hand, given the degeneracy between the two parameters, the constraint on the dark energy equation of state $w_0$ is weakened by about a factor of two. However, we note that there is an additional difference between the two analyses which contributes to these changes: in \citet{AnaMarta2} the Hubble constant was kept fixed at $H_0=70$ km/s/Mpc, while in the present work it is always analytically marginalised. In this sense the present analysis is more conservative, which makes the improvement on the $\zeta$ constraint more significant.

\section{\label{sec:forec}Improving constraints: prospects and limitations}

We now discuss how these constraints may improve with forthcoming cosmological, astrophysical, and local measurements. This also allows us to address the question of what kind of measurements of the fine structure constant we would need in order distinguish between a freezing and a thawing model. For this purpose, and in addition to the standard $\Lambda$CDM model (which has $w_0=-1$ and $\zeta=0$), we consider two fiducial models, with a thawing and freezing behaviour  respectively, with the model parameters
\begin{equation}
\label{eq:fid}
w_0=-0.95\,,  \qquad \zeta=-2.5 ppm\,, \qquad \beta=\pm0.8;
\end{equation}
these are within the current one-sigma constraints presented in the previous section.

We conservatively assume that the sensitivity of the Oklo and atomic clock bounds will be unchanged. Regarding the cosmological data, we assume a future WFIRST-like supernova data set as discussed in \citet{Riess}. These forecasts were computed assuming independent measurements; in \citep{Riess}, the authors claim that any correlations between observations should be moderate, but attempts to contact the authors for clarification have so far been unsuccessful. As for future constraints on the Eotvos parameter, we assume an order-of-magnitude improvement on the current ones --- this may soon be available from the final MICROSCOPE results \citep{Touboul2}.

As for the fine-structure constant $\alpha$, we consider an improved data set of 300 measurements, expected to be put together by facilities such as ESPRESSO \citep{ESPRESSO,ESPRESSO2} which is already operational, and ELT-HIRES \citep{EELT} which is expected to be operational within a decade. We conservatively assume that these measurements span the redshift range $0<z<4$ (which contains almost all currently available measurements), and look at three possible sensitivities, $\sigma=5, 1,$ and $0.1 ppm$. We note that the case of $\sigma=5 ppm$ is comparable to (in fact slightly poorer than) the currently available $\alpha$ data set, but it serves as a proxy for assessing the impact of an improved constraint on the Eotvos parameter while also providing a fair comparison to our two other $\alpha$ data sets.

For each combination of fiducial model and observational sensitivity we generate a mock data set, which we then put through our previously developed statistical analysis pipeline. We have verified that the fiducial model parameters are recovered, but this is subject to caveats that are discussed below.

\begin{table}
\caption{Comparison of two-sigma ($95.4\%$ c.l.) constraints for the current data (cf. Table \protect\ref{table1}) and our three simulated scenarios, assuming that the fiducial model is $\Lambda$CDM. Since the posterior likelihoods are not symmetric, we show the full parameter range containing the two-sigma constraints.}
\label{table2}
\centering
\begin{tabular}{| c | c || c | c | c |}
\hline
Parameter & Current & $\sigma=5$ ppm & $\sigma=1$ ppm & $\sigma=0.1$ ppm  \\
\hline
$w_0$ & 0.14 & 0.16 & 0.15 & 0.11 \\
$\zeta$ (ppm) & 5.3 & 3.0 & 2.3 & 0.5 \\
\hline
\end{tabular}
\end{table}

We start with the scenario where the fiducial model is the standard $\Lambda$CDM. Table \ref{table2} shows how the current upper limits previously shown in Table \ref{table1} would be improved by our simulated data sets. We note that the posterior likelihoods are neither Gaussian nor symmetric, and therefore we list the full range of each parameter which contains the two-sigma constraint. Overall we see that the gains in sensitivity on $w_0$ are comparatively small, while those on $\zeta$ are quite significant. This is a consequence of the fact that constraints on the Eotvos parameter directly correspond to constraints on $\zeta$. Improving the MICROSCOPE constraint by a factor of ten without significantly changing the $\alpha$ data set (our $\sigma=5$ ppm case) is expected to improve the constraint on $\zeta$ by a factor of 1.7 while that on $w_0$ worsens by about 10\%. This worsening is due to the correlation between the two parameters. On the other hand, in the $\sigma=0.1$ ppm case the constraint on $\zeta$ improves by a factor of ten while that on $w_0$ improves by about 20\%. Naturally, $\beta$ is unconstrained in this scenario.

\begin{figure*}
\centering
\textbf{Freezing ($\beta=-0.8$) \hskip5cm Thawing ($\beta=+0.8$)}\par\smallskip
\textbf{$\sigma=5$ ppm}\par\smallskip
\includegraphics[width=9cm]{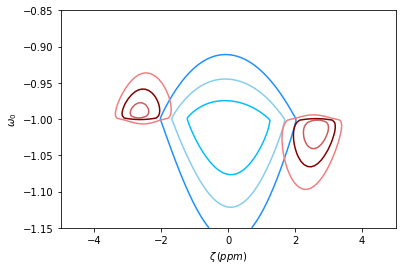}
\includegraphics[width=9cm]{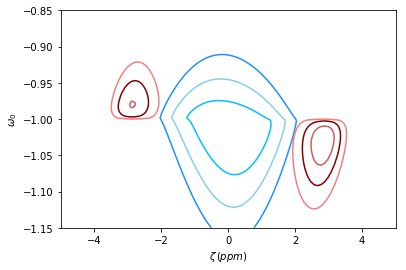}
\textbf{$\sigma=1$ ppm}\par\smallskip
\includegraphics[width=9cm]{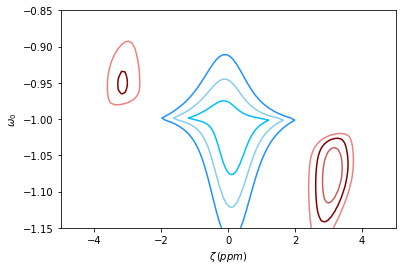}
\includegraphics[width=9cm]{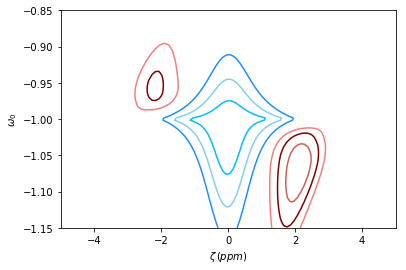}
\textbf{$\sigma=0.1$ ppm}\par\smallskip
\includegraphics[width=9cm]{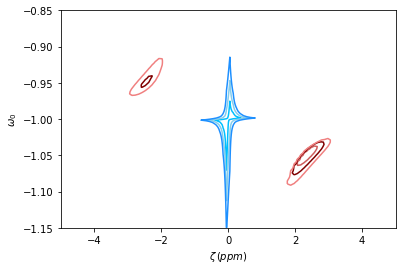}
\includegraphics[width=9cm]{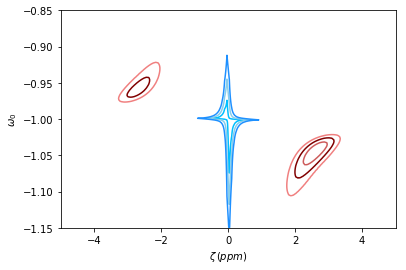}
\caption{Constraints on the $\zeta$-$w_0$ plane for our simulated data sets, with $\beta$ marginalised. The blue contours are for a $\Lambda$CDM fiducial, while the red ones are for a fiducial model with $w_0=-0.95$ and $\zeta=-2.5 ppm$. For the latter fiducial, left and right columns of panels  correspond to $\beta=-0.8$ and $\beta=+0.8,$ respectively, while top, middle, and bottom rows correspond to different sensitivities in the $\alpha$ data set, respectively $\sigma=5, 1,$ and $0.1 ppm$. One, two, and three-sigma confidence levels are shown in all panels.}
\label{fig6}%
\end{figure*}

We now move to the scenario where the fiducial model is $w_0=-0.95$ and $\zeta=-2.5 ppm$, either with $\beta=-0.8$ or $\beta=+0.8$. The results of this analysis are summarised in Figure \ref{fig6}. We see that this model can be distinguished from $\Lambda$CDM, for which corresponding constraints are also depicted in all panels to facilitate the comparison. We also note that the constraints are stronger for freezing $\beta<0$ than for thawing models, and that the canonical fiducial model is always recovered, but a phantom counterpart, corresponding to {\bf $w_0=-1.05$} and $\zeta=+2.5 ppm,$ is also statistically allowed; both of these occur for the reasons mentioned in the previous section. The result is that two-sigma constraints are quite weak and do not allow us to distinguish between the canonical and phantom sectors of the model. The phantom solution is, statistically speaking, slightly preferred over the canonical one, though the difference is not significant; this preference is mainly driven by the cosmological data, as also seen from other data sets \citep{Planck18,Alam}.

A preliminary conclusion is therefore that this parametrisation is too broad if one mathematically allows for its full parameter space. Still it is worth studying three specific cases: those where only one of the two parameters $\zeta$ and $w_0$ differs from the canonical behaviour, and the case where only canonical dark energy equations of state are allowed, and therefore $w_0\ge-1$, which is (arguably) better motivated from a theoretical physics perspective. We separately discuss each of the three in the following sections.

\subsection{Case $w_0=-0.95$ and $\zeta=0$}

This case corresponds to the scenario where the dark energy is not a cosmological constant, being due to the dynamics of some fundamental or effective scalar field, but some unknown mechanism suppresses the coupling of this field to the electromagnetic sector and therefore one has $\zeta=0$. We emphasise that given the current understanding of 4D effective field theory such a suppression would be a form of fine-tuning \citep{Carroll,Dvali,Chiba}, but nevertheless this is a conceptually valid scenario worth discussing.

\begin{figure*}
\centering
\textbf{Freezing ($\beta=-0.8$) \hskip5cm Thawing ($\beta=+0.8$)}\par\smallskip
\textbf{$\sigma=5$ ppm}\par\smallskip
\includegraphics[width=9cm]{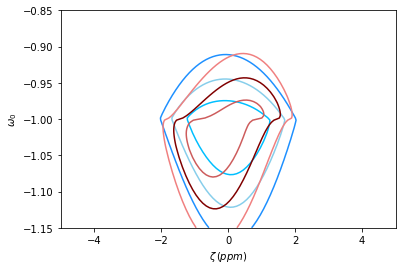}
\includegraphics[width=9cm]{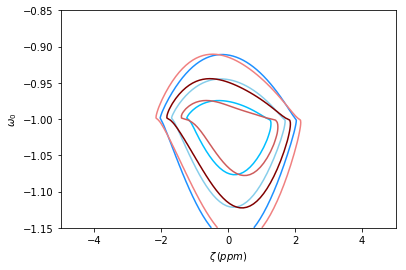}
\textbf{$\sigma=1$ ppm}\par\smallskip
\includegraphics[width=9cm]{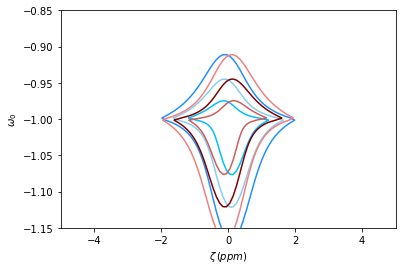}
\includegraphics[width=9cm]{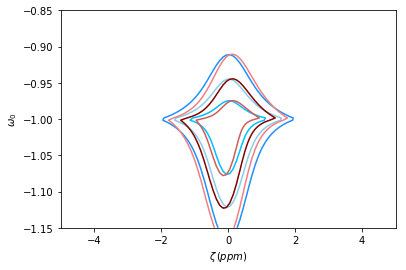}
\textbf{$\sigma=0.1$ ppm}\par\smallskip
\includegraphics[width=9cm]{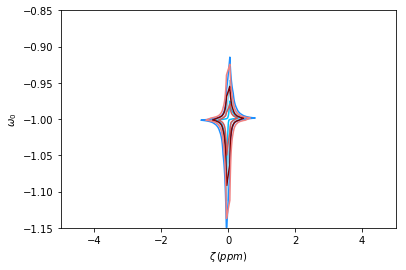}
\includegraphics[width=9cm]{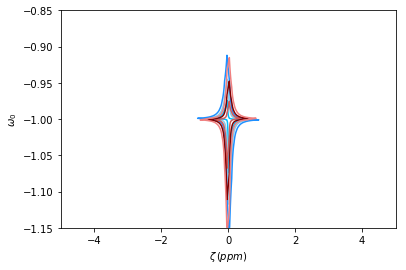}
\caption{Constraints on the $\zeta$-$w_0$ plane for our simulated data sets, with $\beta$ marginalised. The blue contours are for a $\Lambda$CDM fiducial, while the red ones are for a fiducial model with $w_0=-0.95$ and $\zeta=0$. For the latter fiducial, left and right columns of panels  correspond
to $\beta=-0.8$ and $\beta=+0.8,$ respectively, while top, middle, and bottom rows correspond to different sensitivities in the $\alpha$ data set, respectively $\sigma=5, 1,$ and $0.1 ppm$. One, two, and three-sigma confidence levels are shown in all panels.}
\label{fig7}%
\end{figure*}

\begin{table}
\caption{Comparison of two-sigma ($95.4\%$ c.l.) constraints for our three simulated scenarios, assuming that the fiducial model is $w_0=-0.95$ and $\zeta=0$. The first two rows are for freezing models (specifically with $\beta=-0.8$) and the last two for thawing models (specifically with $\beta=+0.8$).  Since the posterior likelihoods are not symmetric, we show the full parameter range containing the two-sigma constraints.}
\label{table3}
\centering
\begin{tabular}{| c | c || c | c | c |}
\hline
Model & Parameter & $\sigma=5$ppm & $\sigma=1$ppm & $\sigma=0.1$ppm  \\
\hline
$\beta=-0.8$ & $w_0$ & 0.18 & 0.15 & 0.10 \\
{ } & $\zeta$ (ppm) & 2.9 & 2.4 & 0.5 \\
\hline
 $\beta=+0.8$ & $w_0$ & 0.17 & 0.15 & 0.11 \\
{ } & $\zeta$ (ppm) & 3.4 & 2.0 & 0.4 \\
\hline
\end{tabular}
\end{table}

The results of our analysis are summarised in Figure \ref{fig7} and Table \ref{table3}; the latter again lists the full range of each parameter which contains the two-sigma constraints. All fiducial models are recovered at the one-sigma confidence level, with the exception of $w_0$ in the $\sigma=0.1$ ppm case, which is only recovered at the two-sigma level. The latter is due to the increased statistical weight of the $\alpha$ measurements, which in this case will be consistent with no variations. Overall the behaviour is the same as was discussed above for the standard model constraints.

\subsection{Case $w_0=-1$ and $\zeta=-2.5 ppm$}

This corresponds to the scenario where the dark energy accelerating the universe is due to a cosmological constant, while a separate physical mechanism leads to variations of the fine-structure constant, and therefore a non-zero $\zeta$. Examples of this scenario include Bekenstein-type models \citep{Bekenstein1,Bekenstein2} and their extensions by \cite{OliveP}. Recent constraints on the former can be found in \citet{LeiteB}, and on the latter in \citet{AlvesO}.

\begin{figure*}
\centering
\textbf{Freezing ($\beta=-0.8$) \hskip5cm Thawing ($\beta=+0.8$)}\par\smallskip
\textbf{$\sigma=5$ ppm}\par\smallskip
\includegraphics[width=9cm]{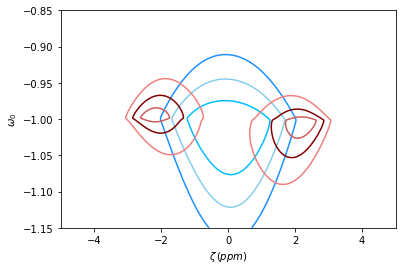}
\includegraphics[width=9cm]{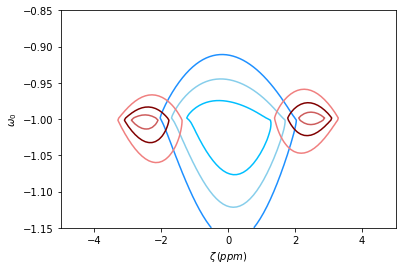}
\textbf{$\sigma=1$ ppm}\par\smallskip
\includegraphics[width=9cm]{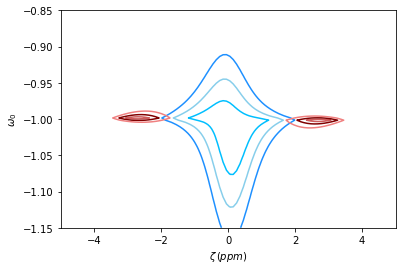}
\includegraphics[width=9cm]{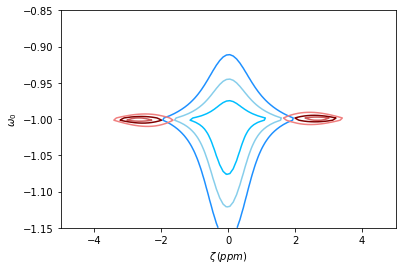}
\textbf{$\sigma=0.1$ ppm}\par\smallskip
\includegraphics[width=9cm]{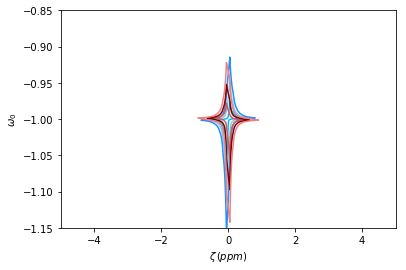}
\includegraphics[width=9cm]{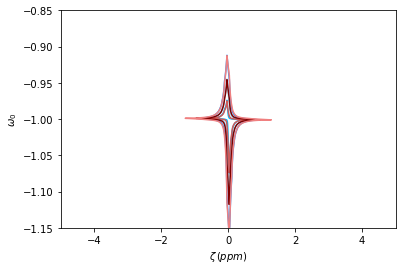}
\caption{Constraints on the $\zeta$-$w_0$ plane for our simulated data sets, with $\beta$ marginalised. The blue contours are for a $\Lambda$CDM fiducial, while the red ones are for a fiducial model with $w_0=-1$ and $\zeta=-2.5 ppm$. For the latter fiducial, left and right columns of panels  correspond
to $\beta=-0.8$ and $\beta=+0.8,$ respectively, while top, middle, and bottom
rows correspond to different sensitivities in the $\alpha$ data set, respectively $\sigma=5, 1,$ and $0.1 ppm$. One, two, and three-sigma confidence levels are shown in all panels.}
\label{fig8}%
\end{figure*}

The results of our analysis are summarised in Figure \ref{fig8}. We find that while the $w_0=-1$ fiducial value is always recovered at one-sigma, this is not necessarily the case for the value of $\zeta$. Indeed, for $\sigma=0.1$ ppm the extremely constraining $\alpha$ data set (which is consistent with no variations, since $w_0=-1$) will lead to a two-sigma constraint $\zeta=0.0_{-0.3}^{+0.2}$, strongly disagreeing with the fiducial. Nevertheless, we note that there is a positive aspect to this. In these models, the non-zero value of $\zeta$ should presumably be detected by one of the local experiments probing the Einstein equivalence principle. In that case, the disagreement between local and astrophysical measurements of $\zeta$ would be an indication that two different mechanisms are behind the dark energy and the Einstein equivalence principle violation. In other words, this would disfavour Class I models and be an indication of an underlying Class II model, as discussed in the introduction and in more detail in \citet{ROPP}.

In the other two cases, $\sigma=5$ and  $\sigma=1$ ppm, the simulated MICROSCOPE bound has a comparatively stronger statistical weight, and the fiducial value of $\zeta$ is now recovered, though with a large uncertainty, as in the general case discussed above. We note that again we see the two branches of the solution, a canonical one with a negative $\zeta$ and a phantom one with a positive $\zeta$. Although statistically there is no strong preference for one or the other, it is interesting to note that in the freezing case ($\beta=-0.8$) the phantom branch is the slightly favoured one, while in the thawing case ($\beta=+0.8$) it is the canonical branch that is slightly favoured.

\subsection{The canonical $w_0\ge-1$ case}

We finally consider the case where we impose that $w_0\ge-1$, which can be seen as a physical choice of prior; phantom models, while a legitimate subject for phenomenological study and statistical analysis, are known to be physically problematic \citep{Phantom1,Phantom2,Phantom3}. Other than this, the analysis is similar to the previous one, in particular assuming uniform priors for all the model parameters.

\begin{figure*}
\centering
\textbf{Freezing ($\beta=-0.8$) \hskip5cm Thawing ($\beta=+0.8$)}\par\smallskip
\textbf{$\sigma=5$ ppm}\par\smallskip
\includegraphics[width=9cm]{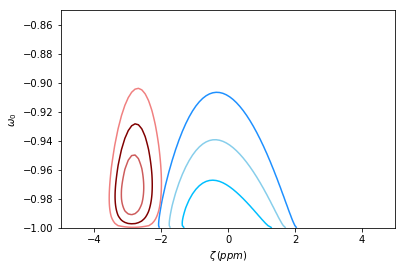}
\includegraphics[width=9cm]{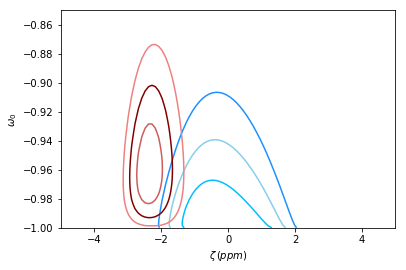}
\textbf{$\sigma=1$ ppm}\par\smallskip
\includegraphics[width=9cm]{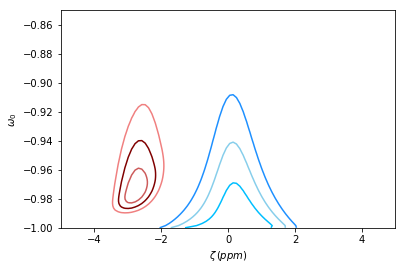}
\includegraphics[width=9cm]{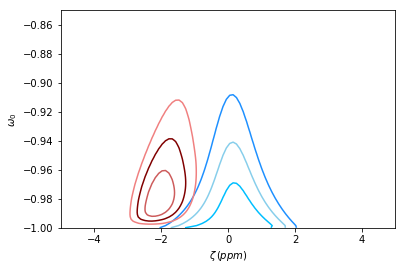}
\textbf{$\sigma=0.1$ ppm}\par\smallskip
\includegraphics[width=9cm]{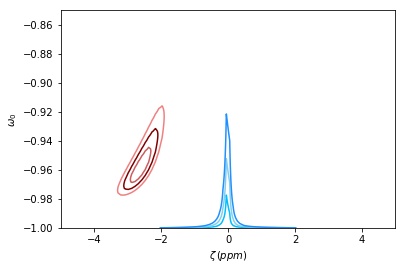}
\includegraphics[width=9cm]{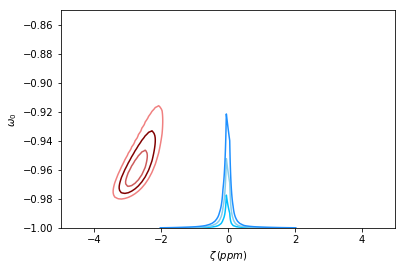}
\caption{Constraints on the $\zeta$-$w_0$ plane for our simulated data sets, with $\beta$ marginalised, with only canonical models ($w_0\ge-1$) allowed. The blue contours are for a $\Lambda$CDM fiducial, while the red ones are for a fiducial model with $w_0=-0.95$ and $\zeta=-2.5 ppm$. For the latter fiducial, left and right columns of panels  correspond
to $\beta=-0.8$ and $\beta=+0.8,$ respectively, while top, middle, and bottom
rows correspond to different sensitivities in the $\alpha$ data set, respectively $\sigma=5, 1,$ and $0.1 ppm$. One, two, and three-sigma confidence levels are shown in all panels.}
\label{fig9}%
\end{figure*}

The results are summarised in Figure \ref{fig9} and Table \ref{table4}. Again we see that the fiducial model is always recovered at one standard deviation. By comparison to the previous cases, the main difference here is that the reduction in the allowed volume of parameter space as the sensitivity of the $\alpha$ measurements improves is now done along the direction of the $\beta$ parameter. This is the reason why the constraints on $w_0$ and $\zeta$ are only improved slightly when going from $\sigma=5$ to $\sigma=0.1$ ppm. On the other hand, with sufficiently precise measurements of $\alpha$ it now becomes possible to constrain $\beta$. This starts to become possible around $\sigma=1$ ppm: for the mock datasets we have generated for this case, for the $\beta=-0.8$ model the constraint is at the level of 1.9 standard deviations (and therefore it narrowly misses the two-sigma threshold used in Table \ref{table4}), while for the $\beta=+0.8$ it is at about 2.1 standard deviations. On the other hand, for $\sigma=0.1$ ppm one cannot only constrain $\beta$ but it even becomes possible to distinguish between freezing and thawing models, at least (under the assumptions used for generating our mock data sets) at the one- to two-sigma level. This case also makes it clear that the degeneracy directions in the $w_0$-$\zeta$ plane change with the sensitivity of the $\alpha$ measurements. This is due to the different relative statistical weight of the astrophysical and local data: for $\sigma=5$ ppm the constraints are clearly dominated by the local data (and in particular by the MICROSCOPE measurement) while for $\sigma=0.1$ ppm the $\alpha$ measurements dominate. In this regard the differences between the freezing and thawing fiducials are comparatively small.

\begin{table}
\caption{Comparison of two-sigma ($95.4\%$ c.l.) constraints for our three simulated scenarios, assuming that the fiducial model is $w_0=-0.95$ and $\zeta=-2.5 ppm$ and allowing only canonical dark energy models. The first three rows are for freezing models (specifically with $\beta=-0.8$) and the last three for thawing models (specifically with $\beta=+0.8$). `Unc.' denotes an unconstrained parameter.}
\label{table4}
\centering
\begin{tabular}{| c | c || c | c | c |}
\hline
Model & Parameter & $\sigma=5$ppm & $\sigma=1$ppm & $\sigma=0.1$ppm  \\
\hline
{ } & $w_0$ & $-0.97_{-0.02}^{+0.05}$ & $-0.97_{-0.01}^{+0.04}$ & $-0.97_{-0.02}^{+0.03}$ \\
 $\beta=-0.8$ & $\zeta$ (ppm) & $-2.9_{-0.5}^{+0.6}$ & $-2.8_{-0.5}^{+0.6}$ & $-2.6\pm0.05$ \\
{ } & $\beta$ & Unc. & Unc. & $-1.0_{-0.6}^{+0.7}$ \\
\hline
{ } & $w_0$ & $-0.96_{-0.03}^{+0.06}$ & $-0.98_{-0.01}^{+0.04}$ & $-0.96_{-0.02}^{+0.03}$ \\
 $\beta=+0.8$ & $\zeta$ (ppm) & $-2.9_{-0.5}^{+0.7}$ & $-2.1_{-0.6}^{+0.8}$ & $-2.7\pm0.05$ \\
{ } & $\beta$ & Unc. & $-0.6_{-1.5}^{+2.5}$ & $0.6_{-0.7}^{+1.0}$ \\
\hline
\end{tabular}
\end{table}

\section{Conclusions}

We used a combination of background cosmology, astrophysical spectroscopy, and   data from laboratory experiments to constrain a phenomenological but physically realistic dynamical dark energy model. This phenomenological model was introduced by \citet{Mukhanov} with the aim of studying inflationary scenarios in the early universe, but is equally applicable to the more recent universe and the dynamical dark energy context. The model is interesting for the latter purpose because it can either have a freezing or thawing behaviour, depending on the value of one of the model parameters, denoted $\beta$.

We allowed for the possibility that the putative dynamical scalar field responsible for the dark energy also couples to the electromagnetic sector, thereby violating the weak equivalence principle and leading to a redshift dependence of the fine-structure constant $\alpha$. It cannot be overemphasised that this is the natural expectation in any realistic (physically motivated) dynamical dark energy scenario, at least if it relies on one or more underlying scalar fields. Suppressing these couplings requires postulating a hitherto unknown symmetry \citep{Carroll}.

The important consequence here is that dynamical dark energy models can---and should---be constrained by a combination of cosmological, astrophysical, and local data. Our analysis takes advantage of recent gains in sensitivity on all three of these fronts to improve the constraints on the key coupling parameter mentioned in the previous paragraph (and denoted $\zeta$ in this work) by more than a factor of two, despite the fact that we consider a more extended parameter space than the one used in previous studies. It is clear that among these more recent data sets the one leading to the tightest constraints is the MICROSCOPE bound on the Eotvos parameter \citep{Touboul}.

We also discuss the improvements expected from near-future facilities, and comment on the practical limitations of this class of parametrisations. Our main aim here is to quantify the impact of the expected gains in sensitivity from MICROSCOPE (whose final results are expected soon, and may be further improved by subsequent facilities) and from high-resolution ultra-stable astrophysical spectrographs such as ESPRESSO (already operational) and the next-generation ELT-HIRES (due to start its Phase B of development in 2020). In order to minimise the number of parameters being varied when generating mock data sets we conservatively assumed that the sensitivity of cosmological data is unchanged (though we always generated mock data compatible with the fiducial model being assumed).

Our results show that if the correct fiducial model is the standard $\Lambda$CDM we may expect the constraints on $\zeta$ to improve by at least one order of magnitude (especially considering that the sensitivity of cosmological data sets will also improve). On the other hand, if the correct fiducial model is not the standard one, then sufficiently sensitive data can distinguish between freezing and thawing models, although in general this requires that one makes the physically reasonable assumption that the relevant parameter space does not include phantom dark energy models (whose dark energy equation of state is more negative than that for a cosmological constant). If phantom models are allowed, $\beta$ is unconstrained.

One of the assumptions underlying our analysis is that the dynamical degree of freedom responsible for the putative $\alpha$ variation is the same one producing the dynamical dark energy---in other words, that it is a Class I model, in the classification of \citet{ROPP}. Nevertheless this hypothesis is testable and falsifiable. This point is illustrated by the scenario where the fiducial model has $w_0=-1$ but $\zeta\neq0$. In this case, sufficiently precise and accurate astrophysical data would find a value consistent with $\zeta=0$, in disagreement with other local measurements which would find $\zeta\neq0$. This would then point to a Class II model, where dark energy could be a cosmological constant while a different dynamical mechanism would be responsible for the Einstein equivalence principle violation. In other words, the dynamical degree of freedom responsible for the latter would not provide the dominant dark energy contribution, and even if this degree of freedom led to a detectable $\alpha$ variation (which would therefore suggest a dark energy equation of state $w_0\neq-1$) the disagreement with cosmological data consistent with $w_0=-1$ would again point to a Class II model. As mentioned above, examples of this scenario include Bekenstein-type models \citep{Bekenstein1,Bekenstein2} and their extensions by \cite{OliveP}.

Our analysis also demonstrates that despite the relevance of having a single parametrisation that can simultaneously account for freezing and thawing models, the Mukhanov parametrisation has two limitations when it is used for the purposes we consider here. The first stems from the fact that freezing models lead (other things being equal) to stronger $\alpha$ variations than thawing models and will therefore be more constrained by null $\alpha$ results. In principle this could be mitigated by choosing priors that compensate for this, but the interpretation of such priors is unclear. The second limitation, as mentioned immediately above, is the fact that a completely generic parameter space (including both canonical and phantom dark energy models) will lead to full degeneracies between models parameters. This can be circumvented by adopting a physical prior that excludes phantom models. The extent to which these features apply to other dynamical dark energy models will be the subject of future work.

\begin{acknowledgements}
This work was financed by FEDER---Fundo Europeu de Desenvolvimento Regional funds through the COMPETE 2020---Operational Programme for Competitiveness and Internationalisation (POCI), and by Portuguese funds through FCT - Funda\c c\~ao para a Ci\^encia e a Tecnologia in the framework of the project POCI-01-0145-FEDER-028987. A.B. and C.S. acknowledge financial support from Programa Joves i Ci\`encia, funded by Fundaci\'o Catalunya-La Pedrera (Spain).
\end{acknowledgements}

\bibliographystyle{aa} 
\bibliography{alpha} 

\end{document}